\documentclass[letterpaper]{article} 
\usepackage{aaai25}  
\usepackage{times}  
\usepackage{helvet}  
\usepackage{subfigure}
\usepackage{mathrsfs}
\usepackage{amssymb}
\usepackage{amsmath}
\usepackage{multirow}
\usepackage{caption}
\usepackage{bm}
\usepackage{xcolor}
\usepackage{diagbox}
\usepackage{courier}  
\usepackage[hyphens]{url}  
\usepackage{graphicx} 
\usepackage{natbib}  
\usepackage{caption} 
\usepackage{algorithm}
\usepackage{algorithmic}
\usepackage{fancyhdr}
\usepackage{adjustbox}
\usepackage{newfloat}
\usepackage{listings}
\usepackage{xr}
\DeclareCaptionStyle{ruled}{labelfont=normalfont,labelsep=colon,strut=off} 
\floatstyle{ruled}

\fancypagestyle{firstpage}{
    \fancyhf{} 
   \fancyfoot[L]{\rule{4cm}{0.5pt}\\[0.5em]
        \small *corresponding author.
    }
}

\newfloat{listing}{tb}{lst}{}
\floatname{listing}{Listing}
\title{BSAFusion: A Bidirectional Stepwise Feature Alignment Network for Unaligned Medical Image Fusion}
\author{
    Huafeng Li\textsuperscript{\rm 1},
    Dayong Su\textsuperscript{\rm 1},
    Qing Cai\textsuperscript{\rm 2}\equalcontrib,
    Yafei Zhang\textsuperscript{\rm 1}\equalcontrib
}
\affiliations{
   
    \textsuperscript{\rm 1}School of Information Engineering and Automation, Kunming University of Science and Technology, \\Kunming 650500, China \\
    \textsuperscript{\rm 2}School of Information Science and Engineering, Ocean University of China, Qingdao 266100, China\\

    hfchina99@163.com, dayongsu@outlook.com, cq@ouc.edu.cn, zyfeimail@163.com
}

\usepackage{bibentry}

\begin{document}

\maketitle

\begin{abstract}
If unaligned multimodal medical images can be simultaneously aligned and fused using a single-stage approach within a unified processing framework, it will not only achieve mutual promotion of dual tasks but also help reduce the complexity of the model. However, the design of this model faces the challenge of incompatible requirements for feature fusion and alignment. To address this challenge, this paper proposes an unaligned medical image fusion method called Bidirectional Stepwise Feature Alignment and Fusion (BSFA-F) strategy. To reduce the negative impact of modality differences on cross-modal feature matching, we incorporate the Modal Discrepancy-Free Feature Representation (MDF-FR) method into BSFA-F. MDF-FR utilizes a Modality Feature Representation Head (MFRH) to integrate the global information of the input image. By injecting the information contained in MFRH of the current image into other modality images, it effectively reduces the impact of modality differences on feature alignment while preserving the complementary information carried by different images. In terms of feature alignment, BSFA-F employs a bidirectional stepwise alignment deformation field prediction strategy based on the path independence of vector displacement between two points. This strategy solves the problem of large spans and inaccurate deformation field prediction in single-step alignment. Finally, Multi-Modal Feature Fusion block achieves the fusion of aligned features. The experimental results across multiple datasets demonstrate the effectiveness of our method.
The source code is available at \url{https://github.com/slrl123/BSAFusion/}.
\end{abstract}

%

\section{Introduction}
\begin{figure}
	\begin{center}
		\includegraphics[width=0.88\linewidth]{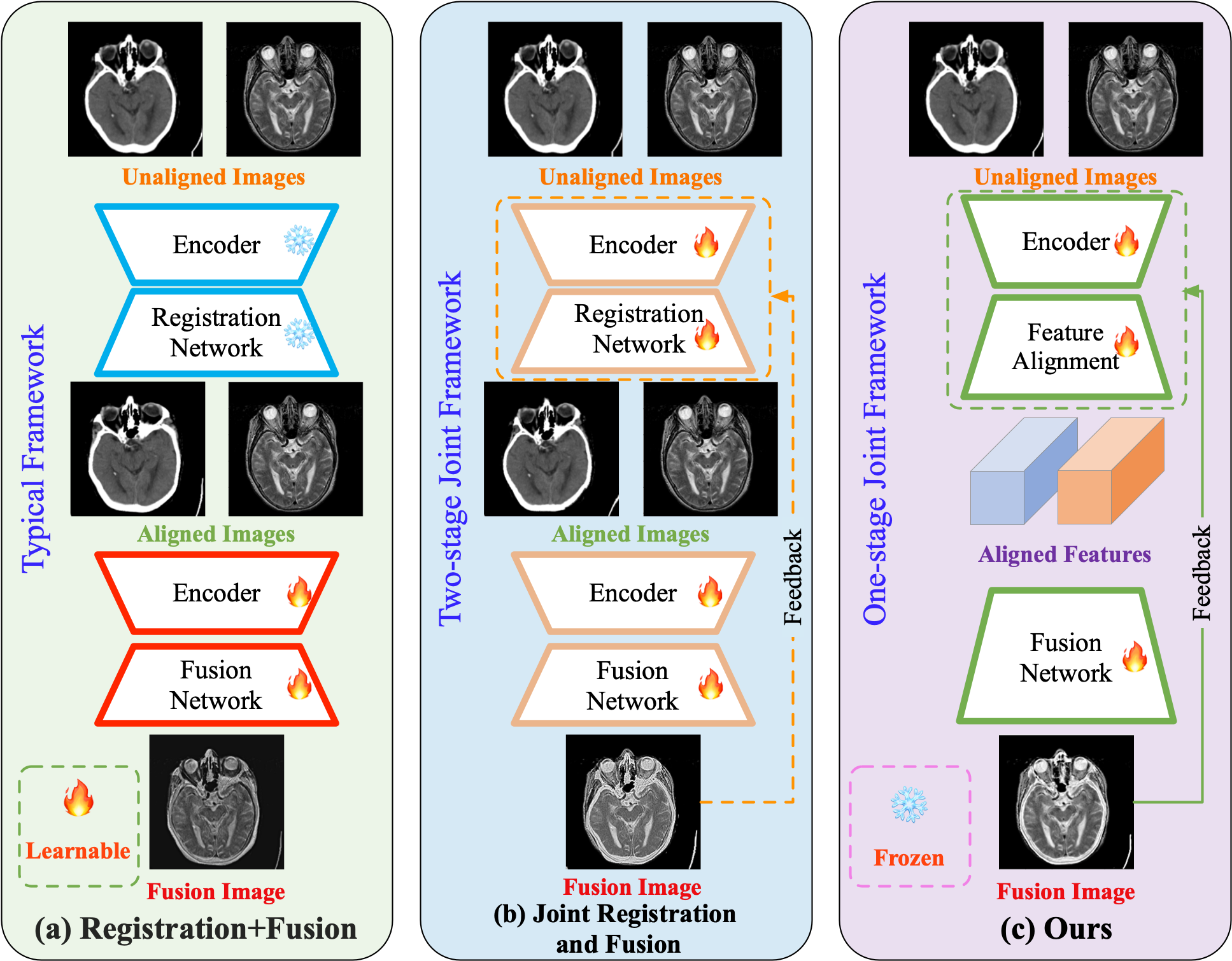}
	\end{center}
	\caption{Paradigm of existing unaligned image fusion methods compared to that of our method.}
	\label{fig1}
\end{figure}

Multimodal medical image fusion (MMIF) involves the integration of medical image data from different imaging modalities (such as CT, MRI, PET, etc.) to create a new image that contains more comprehensive and accurate lesion information. This technology is of great significance in improving diagnostic accuracy, assisting in the development of treatment plans, promoting medical research and education, and optimizing the utilization of medical resources. As a result, it has garnered the attention of researchers, and a multitude of effective fusion algorithms have been proposed \cite{1,2,3}. However, most current methods assume that the source images being fused are strictly aligned at the pixel level. The fusion algorithm can produce the expected results only when this assumption holds true. In real scenarios, however, this assumption is often invalid. To address this issue, registration algorithms are typically used to first align the images to be fused, followed by the fusion process (as shown in Fig.\ref{fig1}(a)). Although this two-stage method is effective, cross-modal image registration still faces numerous challenges due to differences in modalities and inconsistencies in features between images.

In recent years, researchers have begun exploring the integration of multi-source image registration and fusion into a unified framework to address the aforementioned issues. By leveraging the supervision of fusion results, registration performance can be improved. Based on this idea, joint processing frameworks for image registration and fusion have emerged in recent years \cite{5, 6,7,9}. However, these methods are not specifically designed for the registration and fusion of multimodal medical images. Although MURF \cite{11} attempts to integrate multiple types of source image fusion problems into one framework, this multitasking approach often sacrifices performance in single-task image fusion. In view of this, PAMRFuse \cite{10} adopts a similar approach to UMF-CMGR, focusing on the fusion of unregistered multimodal medical images. However, this method depends on an image of one modality to generate a corresponding image in another modality, and its registration performance is often constrained by the quality of the generated image.

In addition, the above methods often adopt a two-stage processing mode (as shown in Fig.\ref{fig1}(b)), with registration preceding fusion. This approach typically necessitates the use of a separate, fully-developed image registration model. This is because registration and fusion have incompatible feature requirements, making it challenging to seamlessly embed both into the fusion process through a shared feature encoder. To tackle this issue, single-stage unaligned fusion methods, such as IVFWSR \cite{13} and RFVIF \cite{14}, have been proposed for infrared-visible image fusion. However, these methods only address feature misalignment caused by rigid transformations and are ineffective in handling elastic transformations. 
In fact, achieving registration and fusion of multimodal medical images using a single-stage processing mode within a joint framework remains challenging. These challenges mainly involve resolving the conflicting requirements of feature extraction for registration and fusion.
Typically, feature fusion expects the features to be complementary, while feature matching demands consistency between corresponding features. To achieve simultaneous feature alignment and fusion within a single-stage processing mode, it is crucial to address the aforementioned issues.

Therefore, this paper proposes a single-stage framework for multimodal medical image registration and fusion (as shown in Fig. \ref{fig1}(c)). Unlike traditional two-stage methods, this approach does not require a separate and complete registration process. Instead, it seamlessly embeds the registration steps into the image fusion process, effectively mitigating the increase in model complexity that would result from introducing multiple independent feature encoders. Technically, we innovatively develop an unaligned medical image fusion method called Bidirectional Stepwise Feature Alignment (BSFA). To effectively mitigate the adverse effects of modality differences on cross-modal feature matching, we integrate the Modality Discrepancy-Free Feature Representation (MDF-FR) method into the BSFA framework. MDF-FR achieves global feature integration by appending a Modality Feature Representation Head (MFRH) to each input image. This method significantly reduces the impact of modality differences and inconsistent multimodal information on feature alignment by injecting the head information of the current image into the features of the other images to be fused. As a result, this design effectively preserves the complementary information carried by different images, ensuring both the integrity and diversity of the data. For feature alignment, we propose a bidirectional stepwise deformation field prediction strategy based on the path independence of vector displacement between two points. This strategy effectively addresses the challenges of large-span and inaccurate deformation field predictions encountered in traditional single-step alignment methods, significantly enhancing both the accuracy and efficiency of feature alignment. Finally, through the Multi-Modal Feature Fusion (MMFF) module, the predicted deformation field is applied to multimodal features, achieving precise alignment and effective fusion of input images at the feature level. Overall, the contributions of this paper can be summarized as follows:

\begin{itemize}
    \item We design a joint implementation framework that integrates feature cross-modal alignment and fusion. By sharing a single feature encoder, it enables the seamless integration of registration and fusion, effectively avoiding the increase in model complexity that would result from introducing additional encoder for registration.
    
    \item We propose a modality discrepancy reduction method. This method achieves global feature integration by appending an MFRH to each input image. By incorporating the feature representation of the current image into the features of the other images to be fused, it effectively mitigates the impact of modality differences on feature alignment.
    
    \item Based on the path independence of vector displacement between two points, a bidirectional stepwise deformation field prediction strategy is proposed. It effectively addresses the challenges of large spans and inaccurate deformation field predictions encountered in traditional single-step alignment methods.
\end{itemize}

\begin{figure*}
	\begin{center}
		\includegraphics[width=0.92\linewidth]{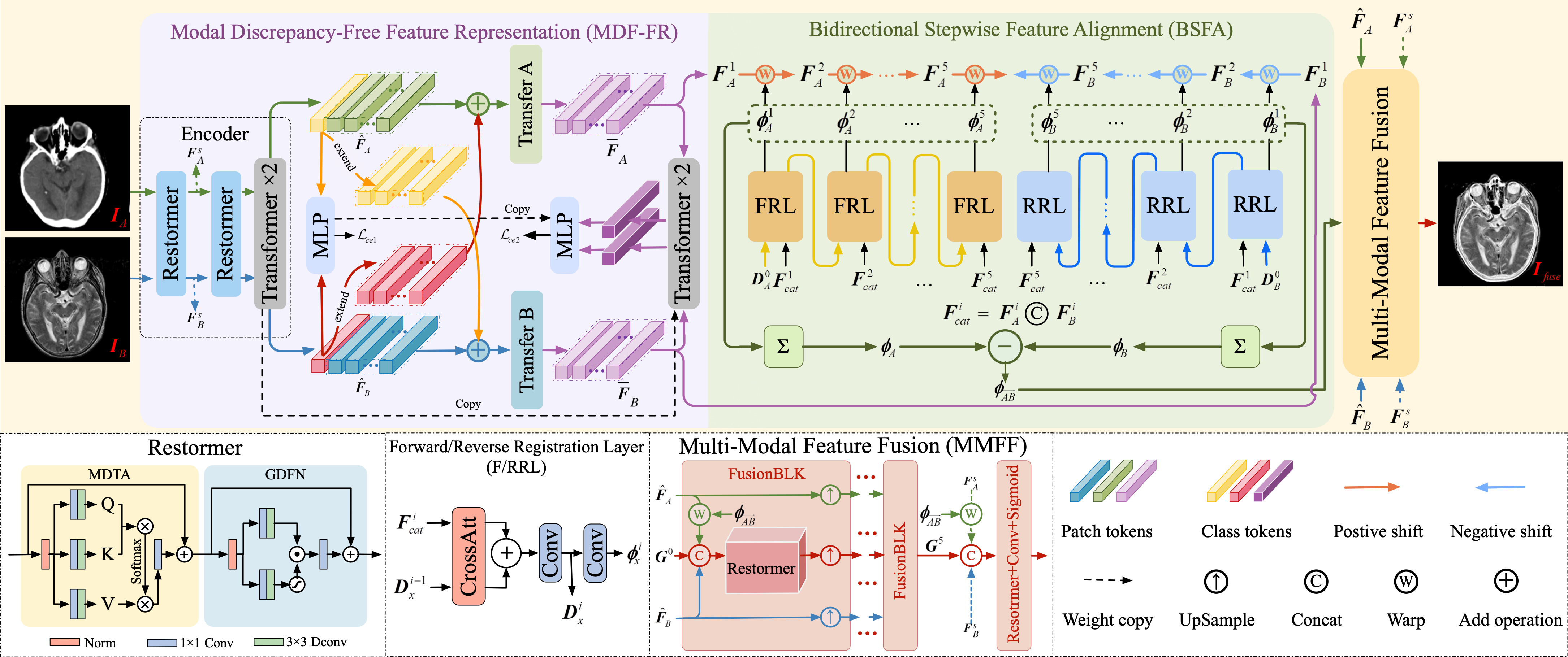}
	\end{center}
	\caption{Overall framework of the proposed method. The unaligned multimodal medical image pairs $\{\bm{I}_{A}, \bm{I}_{B}\}$ are processed through the MDF-FR module, yielding features $\{\bm{F}_{A}^{s}, \bm{F}_{B}^{s}\}$ and $\{\hat{\bm{F}}_{A}, \hat{\bm{F}}_{B}\}$. Additionally, modality-specific feature representation heads, denoted as $\bm{\hat{f}}_{A}$ and $\bm{\hat{f}}_{B}$, are generated. These heads are utilized to minimize the modality disparities between $\{\bm{\hat{F}}_{A}, \bm{\hat{F}}_{B}\}$. Within the BSFA, a progressive deformation field prediction, denoted as $\bm{\phi}_{\overrightarrow{AB}}$, is carried out based on the modality-discrepancy-mitigated features $\{\bm{\bar{F}}_{A}, \bm{\bar{F}}_{B}\}$. Finally, the features $\{\bm{\bar{F}}_{A}, \bm{\bar{F}}_{B}\}$, $\{\bm{F}_{A}^{s}, \bm{F}_{B}^{s}\}$, along with the predicted deformation field $\bm{\phi}_{\overrightarrow{AB}}$, are fed into the MMFF module to generate the final fused result.}
	\label{fig2}
\end{figure*}
\section{Related Work}

For MMIF, deep learning has been widely used due to its ability to effectively extract statistical information from large datasets. Based on the types of feature extraction networks used, existing medical image fusion methods can be classified into CNN-based, Transformer-based, and hybrid methods. Among them, CNN-based methods mainly focus on network architecture design. Commonly used frameworks for MMIF in CNN-based methods include residual connections \cite{15}, skip connections \cite{16}, dense connections \cite{17}, and Network Architecture Search \cite{18}. Additionally, there are dynamic meta-learning method \cite{19} and medical semantic-guided two-branch method \cite{20}. However, these methods are often limited by the shortcomings of CNNs in modeling long-distance dependencies. Transformer \cite{41} has addressed this limitation, resulting in methods such as FATMusic \cite{21} and MATR \cite{2}. Given the complementary strengths of CNN and Transformer in feature extraction \cite{52}, researchers have proposed hybrid methods like DesTrans \cite{22}, DFENet \cite{23} and MRSC-Fusion \cite{24}.

In recent years, various methods have been developed for multimodel image fusion, including MMIF, such as U2Fusion \cite{25}, Cddfuse \cite{26}, DDFM \cite{21},  EMMA \cite{28}, QuadzBayer\cite{50} and HFT\cite{51}. Although these methods are effective, they all assume that the source images to be fused have already been registered. However, in practical applications, this assumption often does not hold true. Therefore, when dealing with unaligned multimodal medical images, these methods cannot be directly applied and require additional image registration models to align the images for fusion. This not only increases model complexity, hindering deployment in computationally constrained environments, but also results in fusion failures if the registration model fails. To address these issues, methods for joint registration and fusion have been developed. Typical examples include ReCoNet \cite{5}, UMF-CMGR \cite{6}, SuperFusion \cite{7}, as well as others like RFNet \cite{9}, MURF \cite{11}, IVFWSR \cite{13}, and MERF \cite{37}. However, these methods are not specifically designed for multimodal medical images and do not exhibit the expected advantages in this domain. Although PAMRFuse \cite{10} is designed for MMIF, its performance is limited by the quality of the generated images. To overcome these challenges, we propose an unaligned MMIF scheme that integrates registration and fusion, allowing the two tasks to complement each other within a single-stage framework.

\section{Proposed Method}
\subsection{Overview}
As shown in Fig. \ref{fig2}, the proposed method consists of three core components: MDF-FR, BSFA, and MMFF. The goal of MDF-FR is to eliminate modality discrepancies between unaligned multimodal medical image pairs ${\bm{I}_{A}, \bm{I}_{B}}$. To address the conflicting requirements of feature alignment and feature fusion, we introduce an MFRH for each input image within MDF-FR. This mechanism reduces the impact of modality discrepancies on feature alignment by injecting MFRH of the current image into the features of other modality image. BSFA predicts the deformation field between features of unaligned images, facilitating subsequent alignment. To tackle challenges posed by large displacements and difficult deformation field predictions inherent in unidirectional methods, BSFA employs a bidirectional, gradually aligned deformation field prediction strategy based on the path independence of vector displacement between two points. Finally, MMFF aligns the features by applying the predicted deformation field and then constructs the fused image based on the aligned features.\vspace{-1mm}

\subsection{Modality Discrepancy-Free Feature Representation}
In MDF-FR, we utilize a network consisting of Restormer and Transformer layers as the encoder for extracting features from unaligned image pairs $\{\bm{I}_{A}, \bm{I}_{B}\}$. The structure of the Restormer is exhibited in Fig. \ref{fig2} \cite{40}. For input images $\bm{I}_{A}$ and $\bm{I}_{B}$, the features output by the first Restormer layer are denoted as $\bm{F}_{A}^s \in \mathbb{R}^{C \times H \times W}$ and $\bm{F}_{B}^s \in \mathbb{R}^{C \times H \times W}$, respectively, where $C$, $H$, and $W$ represent the number of channels, height, and width of the feature maps. Since the shallow features $\bm{F}_{A}^s$ and $\bm{F}_{B}^s$ contain the underlying details of the image, we directly feed them into the multimodal feature fusion layer for feature alignment and fusion to retain more edge details in the fusion results. The features output from the second Restormer layer are then fed into two Transformer layers to extract the features $\bm{\bar{F}}_A$ and $\bm{\bar{F}}_B$, which are used for modality discrepancy elimination and deformation field prediction. The results output by the Transformer layers are denoted as $\bm{\hat{F}}_A = [{\bm{\hat{f}}}_A^1,{\bm{\hat{f}}}_A^2, \cdots, {\bm{\hat{f}}}_A^P] \in \mathbb{R}^{P \times W'}$ and $\bm{\hat{F}}_B = [{\bm{\hat{f}}}_B^1,{\bm{\hat{f}}}_B^2, \cdots, {\bm{\hat{f}}}_B^P] \in \mathbb{R}^{P \times W'}$, along with the modality feature representation heads ${{\bm{\hat{f}}}_A} \in \mathbb{R}^{1 \times W'}$ and ${{\bm{\hat{f}}}_B} \in \mathbb{R}^{1 \times W'}$, where $P$ is the total number of patches that the feature output by the second Restormer layer is divided into, and $W'$ is the length of the vector. In the proposed method, ${{\bm{\hat{f}}}_A}$ and ${{\bm{\hat{f}}}_B}$ are used to describe the modality categories of the input images. To ensure that ${{\bm{\hat{f}}}_A}$ and ${{\bm{\hat{f}}}_B}$ contain the modal information of the input images, we feed them into an MLP, with the output aiming to minimize the cross-entropy loss defined in Eq. (1):
\begin{equation}\small
	\begin{aligned}
		{\cal L}_{ce1} &= CE\left( \bm{y}_A, [0,1] \right) + CE\left( \bm{y}_B, [1,0] \right)
	\end{aligned}
\end{equation}
where $CE$ represents cross entropy, and $\bm{y}_A$ and $\bm{y}_B$ represent the results predicted by the MLP.

Due to the significant modality discrepancies between $\bm{\hat {F}}_{A}$ and $\bm{\hat {F}}_{B}$, cross-modal matching and deformation field prediction based on these features face substantial challenges. To address this issue, existing methods typically extract shared modality features directly from $\bm{\hat {F}}_{A}$ and $\bm{\hat {F}}_{B}$ for deformation field prediction. Although this approach is effective, it may lead to the loss of non-shared or modality-specific information, thereby reducing the expressive power of the features. In contrast, if we directly inject the corresponding modality information into $\bm{\hat {F}}_{A}$ and $\bm{\hat {F}}_{B}$, we can not only mitigate the impact of modality discrepancies on deformation field prediction but also prevent the loss of non-shared information caused by extracting shared information. The modal feature representation heads obtained by minimizing the loss in Eq.(1), which represent global features $\bm{\hat f}_{A}$ and $\bm{\hat f}_{B}$, are not affected by misaligned source images. Therefore, they can be directly injected into $\bm{\hat {F}}_{A}$ and $\bm{\hat {F}}_{B}$ to reduce the modality discrepancies between the two:
\begin{equation}\footnotesize
	\begin{aligned}
		{{\bm{\tilde {F}}}_A} = [{\bm{\hat f}}_A^1 + {{\bm{\hat f}}_B},{\bm{\hat f}}_A^2 + {{\bm{\hat f}}_B}, \cdots {\bm{\hat f}}_A^P + {{\bm{\hat f}}_B}]\\
		{{\bm{\tilde {F}}}_B} = [{\bm{\hat f}}_B^1 + {{\bm{\hat f}}_A},{\bm{\hat f}}_B^2 + {{\bm{\hat f}}_A}, \cdots {\bm{\hat f}}_B^P + {{\bm{\hat f}}_A}]
	\end{aligned}
\end{equation}

To ensure that the features processed by Eq. (2) effectively eliminate differences between modalities, we use two Transfer blocks, namely TransferA and TransferB. Each block is composed of two Transformer layers, and the parameters are not shared between the blocks, allowing for further extraction of features $\bm{\bar{F}}_A$ and $\bm{\bar{F}}_B$ necessary for predicting the deformation field. Additionally, to determine whether the features output by TransferA and TransferB exhibit modality specificity, we replicate the Transformer layer in the encoder and the MLP behind the encoder. We then sequentially pass the features $\bm{\bar{F}}_A$ and $\bm{\bar{F}}_B$ through the Transformer layer and the MLP, respectively, to identify the modality category of features $\bm{\bar{F}}_A$ and $\bm{\bar{F}}_B$. To achieve this, we utilize the cross-entropy loss function in Eq. (3) to update the parameters in TransferA and TransferB:
\begin{equation}\small
	\begin{aligned}
		{{\cal L}_{ce2}} = CE\left( {\bm{y}_A^*,[0.5,0.5]} \right) + CE\left( {\bm{y}_B^*,[0.5,0.5]} \right)
	\end{aligned}
\end{equation}
where $\bm{y}_A^*$ and $\bm{y}_B^*$ are the predicted results of the MLP for the modal categories of $\bm{\bar{F}}_A$ and $\bm{\bar{F}}_B$. In this process, we introduce TransferA and TransferB to further extract features from ${\bm{\tilde {F}}}_A$ and ${\bm{\tilde {F}}}_B$ that are helpful for deformation field prediction. After processing with TransferA and TransferB, we directly reuse the Transformer layer and the MLP to ensure that the features $\bm{\bar{F}}_A$ and $\bm{\bar{F}}_B$ used for deformation field prediction no longer exhibit modality discrepancies.

\subsection{Feature Alignment}
\begin{figure}
	\begin{center}
		\includegraphics[width=0.9\linewidth]{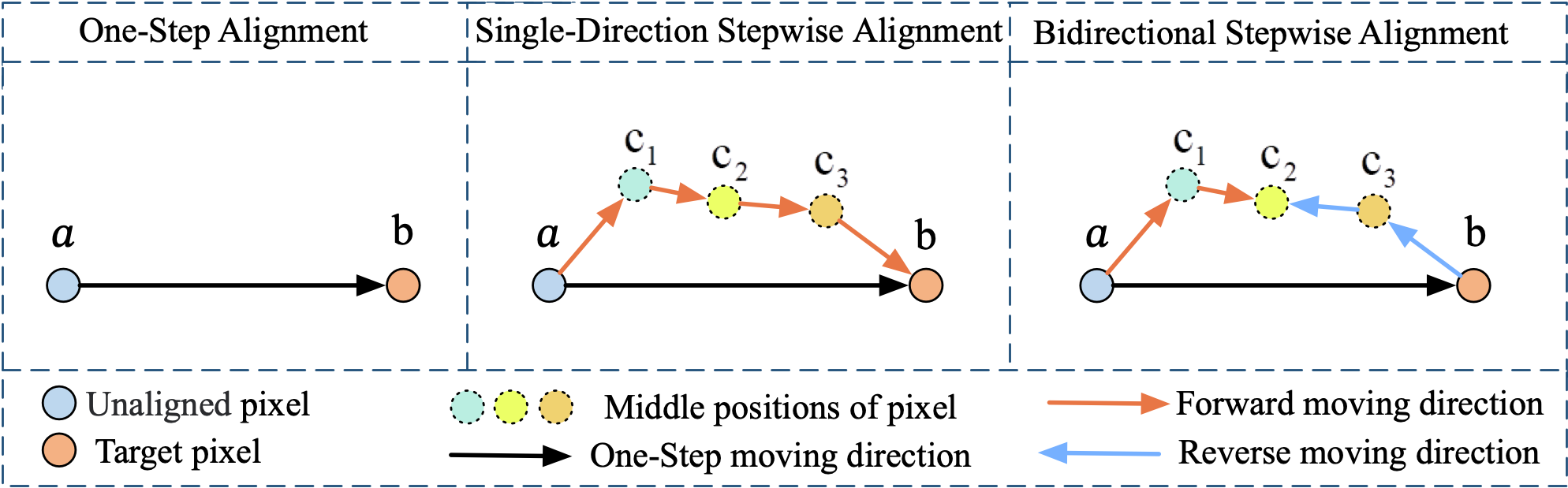}\vspace{-1mm}
	\end{center}
	\caption{Schematic of different alignment methods.}\vspace{-4mm}
	\label{fig3}
\end{figure}

\textbf{Path independence of vector displacement}:
Since ${\bm{\bar {F}}}_A$ and ${\bm{\bar {F}}}_B$ have effectively eliminated the modality discrepancies under the constraint of Eq. (3), it is suitable to predict the deformation field between the features of the input images ${{\bm{I}}_A}$ and ${{\bm{I}}_B}$ using ${{\bm{\bar {F}}}_A}$ and ${{\bm{\bar {F}}}_B}$. However, when the offset between ${{\bm{I}}_A}$ and ${{\bm{I}}_B}$ is large, directly predicting the deformation field between them becomes challenging. As shown in Fig. \ref{fig3}, during the pixel alignment process, pixel $a$ moves to the location of pixel $b$. This can be achieved by applying the deformation field between pixels $a$ and $b$ to $a$. The deformation field between pixels $a$ and $b$, which includes the direction and distance of spatial movement from the location of $a$ to that of $b$, can be regarded as a vector, denoted as $\overrightarrow{ab}$. From Fig. \ref{fig3}, it can be seen that the alignment between the locations of $a$ and $b$ can be achieved in a single step, directly moving from the location of $a$ to that of $b$, or by passing through intermediate points $c_1$, $c_2$, and $c_3$ to reach the location of $b$.

Due to
\begin{equation}\small
	\begin{aligned}
    \overrightarrow {ab}  = \overrightarrow {a{c_1}}  + \overrightarrow {{c_1}{c_2}}  + \overrightarrow {{c_2}{c_3}}  + \overrightarrow {{c_3}b} 
	\end{aligned},
\end{equation}
this indicates that the deformation field from point $a$ to point $b$ can be accumulated from the deformation fields from $a$ to $c_1$, $c_1$ to $c_2$, $c_2$ to $c_3$, and $c_3$ to $b$. This demonstrates that the vector from point $a$ to point $b$ is independent of the path taken from point $a$ to point $b$. In this paper, we refer to this characteristic as the path independence of vector displacement between two points. The deformation field prediction method based on this theory is called the unidirectional progressive prediction method.

As our goal is to align points $a$ and $b$ in their spatial locations, this can be achieved by simultaneously moving both $a$ and $b$ towards an intermediate location. The bidirectional progressive alignment, as shown in Fig.\ref{fig2}, allows point $a$ to reach the intermediate point $c_2$ through point $c_1$, while point $b$ can reach $c_2$ through point $c_3$ in the opposite direction, thereby achieving feature alignment at the location of point $c_2$. At this point, the deformation field used by pixel $a$ to move to the position of pixel $b$ can be expressed as a vector:
\begin{equation}\small
	\begin{aligned}
		\overrightarrow{ab} = \overrightarrow{a{c_2}} - \overrightarrow{b{c_2}}
		= (\overrightarrow{a{c_1}} + \overrightarrow{{c_1}{c_2}}) - (\overrightarrow{b{c_3}} + \overrightarrow{{c_3}{c_2}})
	\end{aligned}
\end{equation}
This bidirectional alignment method effectively captures the interrelationships between images and reduces cumulative errors. Furthermore, if alignment in one direction encounters issues, alignment in the opposite direction can compensate, thereby enhancing the overall robustness of the alignment process. Based on these considerations, this paper proposes the Bidirectional Stepwise Feature Alignment method.

\textbf{Bidirectional Stepwise Feature Alignment}: As shown in Fig. \ref{fig2}, the proposed BSFA predicts the deformation fields of the input image features ${\bm{\bar {F}}}_A$ and ${\bm{\bar {F}}}_B$ from two directions. Both the forward and reverse predictions of BSFA involve five layers of deformation field prediction operations, corresponding to the insertion of five intermediate nodes between the two input source images, achieving spatial alignment of their features by the fifth layer. In our method, the modules responsible for predicting the forward and reverse deformation fields are referred to as the Forward Registration Layer (FRL) and the Reverse Registration Layer (RRL), respectively, as illustrated in Fig. \ref{fig2}. For the initial FRL, its inputs are $\bm{F}_{cat}^1 = \text{concat}( {\bm{F}_A^1, \bm{F}_B^1})$ and $\bm{D}_A^0 \in {\mathbb{R}^{W' \times \frac{H}{{{2^{K - 1}}}} \times \frac{W}{{{2^{K - 1}}}}}}$, and the outputs are the deformation field $\bm{\phi}_A^1$ and $\bm{D}_A^1$. Here, $\bm{F}_A^1 \in {\mathbb{R}^{W' \times \frac{H}{{{2^{K - 1}}}} \times \frac{W}{{{2^{K - 1}}}}}}$ and $\bm{F}_B^1 \in {\mathbb{R}^{W' \times \frac{H}{{{2^{K - 1}}}} \times \frac{W}{{{2^{K - 1}}}}}}$ denote the reshaped versions of ${\bm{\bar {F}}}_A$ and ${\bm{\bar {F}}}_B$, $\bm{D}_A^0$ is initialized to $\bm{F}_A^1$, and $K$ represents the number of layers in both FRL and RRL. The first RRL takes $\bm{F}_{cat}^1$ and $\bm{D}_B^0$ as inputs, and its outputs include the deformation field $\bm{\phi}_B^1$ and $\bm{D}_B^1$, with $\bm{D}_B^0$ initialized to $\bm{F}_B^1$. In the $i$-th FRL and $i$-th RRL, their inputs are $\left\{ \bm{F}_{cat}^i, \bm{D}_A^{i - 1} \right\}$ and $\left\{ \bm{F}_{cat}^i, \bm{D}_B^{i - 1} \right\}$, respectively, where 
\begin{equation}\small
	\begin{aligned}
		\bm{F}_{\text{cat}}^i &= \text{concat}\left( \bm{F}_A^i, \bm{F}_B^i \right) \\
		\bm{F}_j^i &= \uparrow_{\times 2} \left( \bm{W} \left( \bm{F}_j^{i - 1} \mid \bm{\phi}_j^{i - 1} \right) \right), \quad (j = A, B)
	\end{aligned}
\end{equation}
In Eq. (6), $\bm{W}$ represents the Warp operation \cite{29}, which adjusts the spatial position of pixels according to the deformation field $\bm{\phi}_A^{i - 1}$. The symbol ``$\uparrow_{\times 2}$'' denotes a $2\times$ upsampling operation.

After correcting the input image features $\bm{F}_a^i$ and $\bm{F}_b^i$ using the predicted deformation fields in both directions, cross-modal alignment of the features is achieved at an intermediate location. However, the progressive alignment process causes features to gradually move from their original positions. Directly fusing the intermediate aligned features does not allow the input source images to guide the fusion process effectively, which can hinder the improvement of fusion quality. Based on the path independence of vector displacement between two points mentioned earlier, we can construct a transformation that directly aligns the features of source image $\bm{I}_a$ with those of source image $\bm{I}_b$, using the predicted deformation field at each stage. According to the principle in Eq. (5), the deformation field that achieves the alignment of $\bm{I}_a$ and $\bm{I}_b$ features can be expressed as:
\begin{equation}\footnotesize
	\begin{aligned}
{\bm{\phi} _A} &= \mathop \sum \limits_{i = 1}^K { \uparrow _{{2^{K - i}}}}\left( {{2^{K - i}}\bm{\phi} _A^i} \right),
{\bm{\phi} _B} = \mathop \sum \limits_{i = 1}^K { \uparrow _{{2^{K - i}}}}\left( {{2^{K - i}}\bm{\phi} _B^i} \right)\\
{\bm{\phi} _{\overrightarrow {AB} }} &= {\bm{\phi} _A} - {\bm{\phi} _B}
	\end{aligned}
\end{equation}
To ensure the quality of the deformation field, we introduce smoothing loss ${{\cal L}_{smooth}}$:
\begin{equation}\small
	\begin{aligned}
		{\cal L}_{smooth} = \sum_{i=1}^{K} 10^{i - K} \left( \|\nabla \bm{\phi}_A^i\|_2 + \|\nabla \bm{\phi}_B^i\|_2 \right)
	\end{aligned}
\end{equation}
and consistency loss ${{\cal L}_{consis}}$ for model updating:
\begin{equation}\scriptsize
	\begin{aligned}
		{\cal L}_{consis} = {\cal L}_{ssim}\left( \bm{I}'_A, \mathop{\bm W}\nolimits(\bm{I}_A, \bm{\phi}_{\overrightarrow{AB}}) \right) + \left\| \bm{I}'_A - \mathop{\bm W}\nolimits(\bm{I}_A, \bm{\phi}_{\overrightarrow{AB}}) \right\|_1
	\end{aligned}
\end{equation}
where ${\cal L}_{ssim}$ represents structural similarity (SSIM) loss, ${\bm{I}'_A}$ is the label image after ${\bm{I}_A}$ is strictly aligned to ${\bm{I}_B}$ at the pixel level.

\subsection{Multimodal Feature Fusion}
As shown in Fig. \ref{fig2}, after obtaining ${\bm{\phi}_{\overrightarrow{AB}}}$, we send it along with features $\bm{F}_A^s$ and $\bm{F}_B^s$, as well as features $\bm{\hat{F}}_A$ and $\bm{\hat{F}}_B$, to FusionBLK for feature fusion processing. We first transform the features $\bm{\hat{F}}_A \in \mathbb{R}^{P \times W'}$ and $\bm{\hat{F}}_B \in \mathbb{R}^{P \times W'}$ into $\bm{\hat{F}}_A^r \in \mathbb{R}^{W' \times \frac{H}{2(J-1)} \times \frac{W}{2(J-1)}}$ and $\bm{\hat{F}}_B^r \in \mathbb{R}^{W' \times \frac{H}{2(J-1)} \times \frac{W}{2(J-1)}}$, and send them to FusionBLK, where $J$ is the number of FusionBLKs. For the $i$-th FusionBLK, its inputs are $\bm{\hat{F}}_1^r$, $\bm{\hat{F}}_B^r$, and $\bm{G}^{i-1}$, and the output is $\bm{G}^i$:
\begin{equation}\small
	\begin{aligned}
		\bm{G}^i = & \ \uparrow_{\times 2} \bm{E}_r \bigg( \text{concat} \left( \bm{W} \left( \uparrow_{ \times 2^{i - 1}} \left(\bm{\hat{F}}_A^r \right) \middle| \downarrow_{ \times 2^{i - J}} \left( \frac{\bm{\phi}_{\overrightarrow{AB}}}{2^{J - i}} \right) \right), \right. \\
		& \left. \uparrow_{ \times 2^{i - 1}} \left( \bm{\hat{F}}_B^r \right), \bm{G}^{i - 1} \right) \bigg)
	\end{aligned}
\end{equation}
where $\bm{E}_r$ represents the encoder composed of Restormers, and $\bm{G}^0$ is the zero matrix when $i=1$. The output of the last layer, i.e., the $J$-th FusionBLK, is denoted as $\bm{G}^J$. Subsequently, we concatenate $\bm{G}^J$ and $\bm{F}_B^s$ with the corrected result $\bm{\tilde{F}}_A^s = \bm{W}(\bm{F}_A^s\mid\bm{\phi}_{\overrightarrow{AB}})$ of $\bm{F}_A^s$, and send the concatenated result to a reconstruction layer composed of a Restormer, a convolutional layer, and a sigmoid activation function to obtain the fused image $\bm{I}_{fuse}$. The entire fusion process is depicted as ``MMFF'' in Fig.\ref{fig2}, where it represents the sequence of steps involved in the fusion.
\begin{figure*}[t!]
	\begin{center}
		\includegraphics[width=0.90\linewidth]{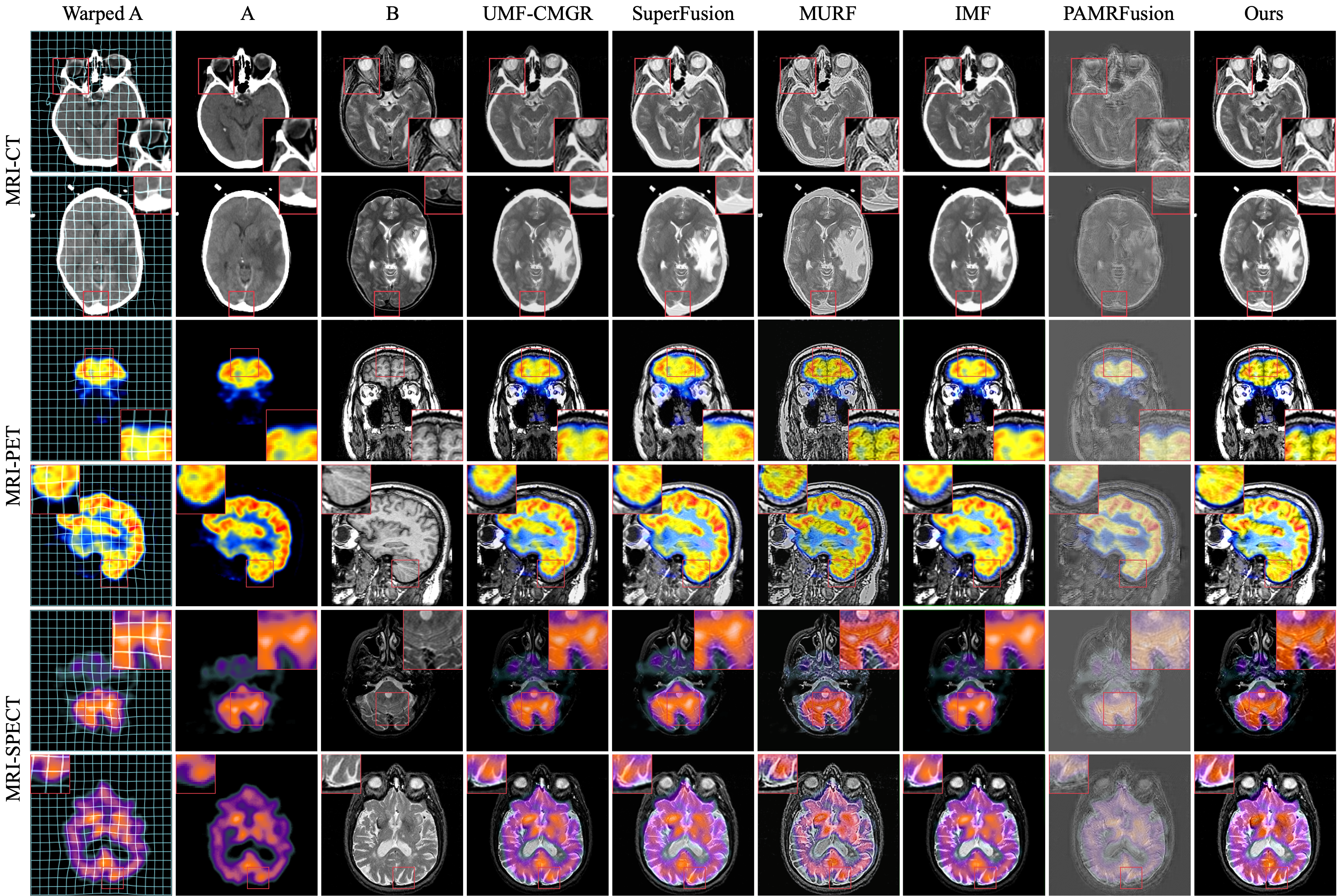}
	\end{center}
	\caption{Visual Comparison of Fusion Results: Joint Registration and Fusion Method vs. Our Method. The first column shows the deformed image to be fused, the second column displays the corresponding label, and the third column presents the MRI image to be fused. Columns 4 to 9 show the results obtained by different fusion methods.}
	\label{fig4}
\end{figure*}

To ensure structural consistency between the fused image and the source images, we use structure loss ${\cal L}_{struct}$ to optimize the network parameters:
\begin{equation}\small
	{\cal L}_{struct} = {\cal L}_{ssim}\left( {\bm{I}_{fuse}, \bm{\tilde{I}}_A} \right) + \mu \, {\cal L}_{ssim}\left( {\bm{I}_{fuse}, \bm{I}_B} \right)
\end{equation}
where $\bm{\tilde{I}}_A = \bm{W}(\bm{I}_A \mid \bm{\phi}_{\overrightarrow{AB}})$, and $\mu$ is a hyperparameter used to adjust the influence of the two SSIM losses in ${\cal L}_{struct}$. To ensure good contrast in the fusion results, we introduce a pixel intensity loss:
\begin{equation}\small
	{\cal L}_{inten} = \left\| {\bm{I}_{fuse} - \max(\bm{\tilde{I}}_A, \bm{I}_B)} \right\|_1
\end{equation}
At the same time, gradient loss is also introduced to prevent the loss of edge details in the source image:
\begin{equation}\small
		\begin{aligned}
 	   {{\cal L}_{grad}} = \left\| {\nabla \bm{I}_{fuse} - \max(\nabla \bm{\tilde{I}}_A, \nabla \bm{I}_B)} \right\|_1
 	   	\end{aligned}
\end{equation}
Therefore, the total loss of this approach is:
\begin{equation}\small
	\begin{aligned}
        {\cal L}_{total} &= {{\cal L}_{ce1}} + {{\cal L}_{ce2}} + {{\cal L}_{consis}} + {{\cal L}_{smooth}} \\&+ {{\cal L}_{struct}} + {{\cal L}_{grad}} + \lambda {{\cal L}_{inten}}	
 	   	\end{aligned}
\end{equation}
where $\lambda$ is a hyperparameter used to adjust the contribution of ${\cal L}_{inten}$ to the total loss.

\section{Experiments}
\subsection{Experimental Setup} 
\textbf{Dataset and Implementation Details}: We follow the protocols of existing methods \cite{30,20} and train the model using CT-MRI, PET-MRI, and SPECT-MRI datasets from Harvard\footnote{http://www.med.harvard.edu/aanlib/}. These datasets consist of 144, 194, and 261 strictly registered image pairs, respectively, each with a size of 256$\times$256. To simulate misaligned image pairs as collected in real-world scenarios, we designate the MRI images as the reference and apply a mixture of rigid and non-rigid deformations to the non-MRI images, thereby creating the required training set. Additionally, the same deformations are applied to 20, 55, and 77 strictly registered image pairs to construct an unaligned test set. To augment the data, we apply these mixed deformations randomly in each epoch, along with random rotations and flips to increase the diversity of the training samples.

During the training process, we adopt an end-to-end approach, training for 3,000 epochs on each dataset with a batch size of 32. We use the Adam optimizer \cite{39} to update model parameters, starting with an initial learning rate of $5\times10^{-5}$. The learning rate is dynamically adjusted using a Cosine Annealing Learning Rate (LR) scheduler \cite{38}, decreasing to $5\times10^{-7}$ over time. Two hyperparameters are set in the loss function: $\lambda$ is set to 0.5, and $\mu$ is updated after each epoch based on the fusion results and the SSIM between the two source images, calculated as $\mu  = \sum\limits_{n = 1}^{N} {{\cal L} _{ssim}^{(n)}( {{\bm{I}_{fuse}},{{\bm{I}_B}}} )/\sum\limits_{n = 1}^{N} {{\cal L}_{ssim}^{(n)}( {{\bm{I}_{fuse}},{{\bm{\tilde I}_{A}}}})} } $, where $N$ is the number of training samples in each epoch. The proposed method is implemented using the PyTorch framework and trained on a single NVIDIA GeForce RTX 4090 GPU.

\textbf{Evaluation Metrics}: To objectively evaluate the performance of fusion methods, we selected five commonly used image quality metrics: Gradient-based Fusion Performance ($Q_{AB/F}$) \cite{32}, Chen-Varshney Metric ($Q_{CV}$)  \cite{33}, Visual Information Fidelity ($Q_{VIF}$) \cite{34}, Structure-based Metric ($Q_S$) \cite{35}, and Structural Similarity Index Measure ($Q_{SSIM}$) \cite{36}. Among these evaluation metrics, a lower value of $Q_{CV}$ indicates better quality of the fused image, while higher values of the other metrics indicate better fusion quality.\vspace{-1mm}
\begin{figure}[t!]
	\centering
	\includegraphics[width=0.43\textwidth]{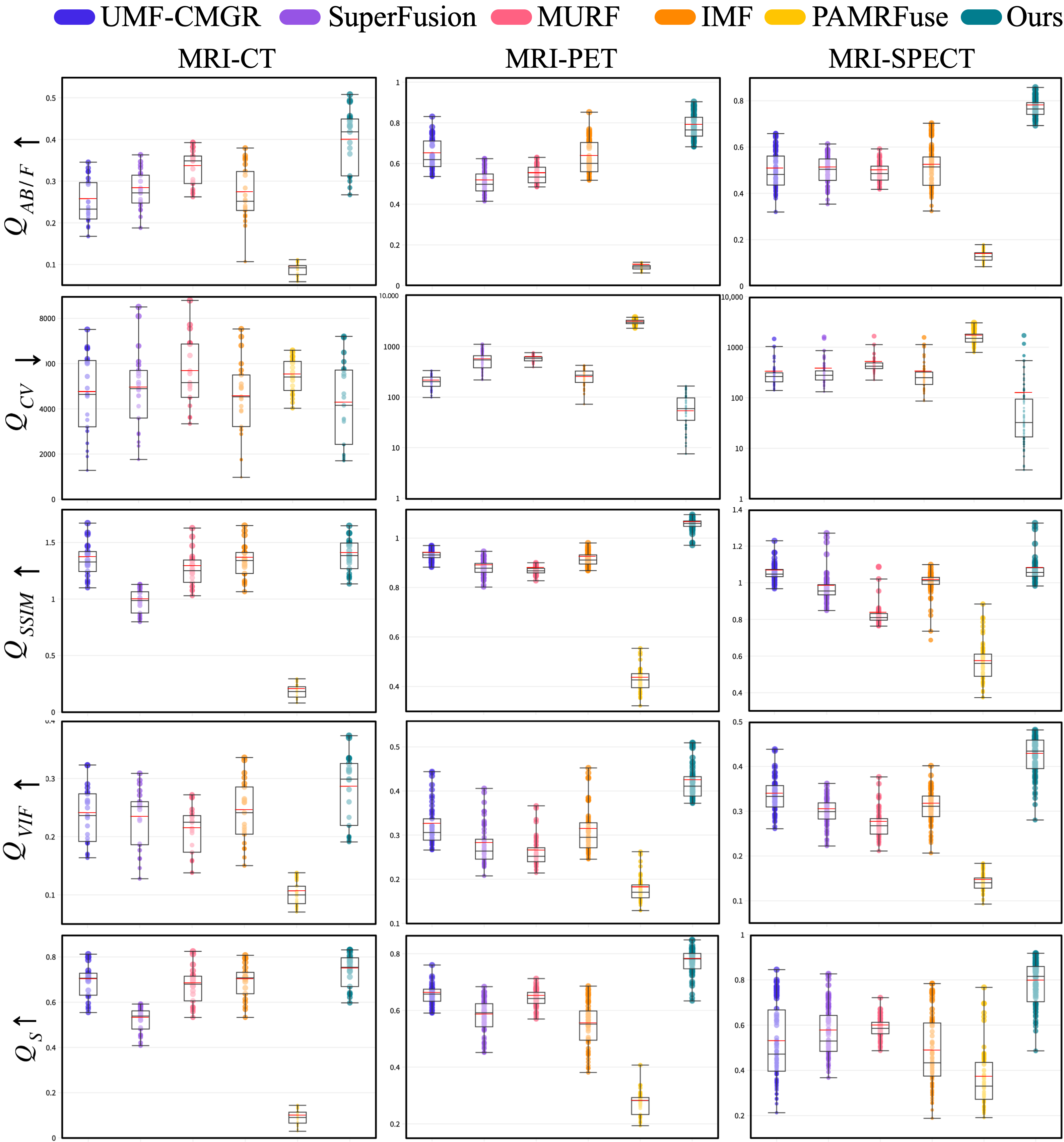}
	\caption{Comparison of objective evaluation results: Joint registration and fusion vs. the proposed method. The black line denotes the median, and the red line denotes the mean.}
	\label{fig5}
\end{figure}
\subsection{Comparison With the State-of-the-art Methods}\vspace{-1mm}
The common approach to solving the problem of unaligned multi-source image fusion is to first perform registration on the images to be fused, and then fuse them. We refer to this method as ``Registration+Fusion." In addition to conventional methods, a two-stage approach has been developed that combines registration and fusion into a single process, referred to as ``Joint Registration and Fusion." To verify the superiority of our method, we compare it with these two approaches. Since our method is most closely related to ``Joint Registration and Fusion," we focus on the comparison with this method here. Due to space limitations, the comparison between our method and the ``Registration+Fusion" method is included in the supplementary materials. 

Specifically, we compared the performance of our method with five joint registration and fusion methods: UMF-CMGR, SuperFusion, MURF, IMF \cite{49}, and PAMRFuse. The first four methods are specifically designed for the registration and fusion of multimodal images and are applicable to MMIF. PAMRFuse, on the other hand, is a method specifically proposed for the fusion of unregistered medical images. Fig. \ref{fig4} presents a visual comparison of the fusion results generated by different methods. It is evident that our proposed method demonstrates significant advantages in feature alignment, contrast preservation, and detail retention. This indicates that, compared to existing two-stage joint processing frameworks, our method exhibits stronger performance, primarily due to its ability to seamlessly integrate registration and fusion tasks into a unified process. Additionally, we created box plots of test metrics for each method to visually analyze performance differences. As shown in Fig. \ref{fig5}, our method achieved the best mean performance across all metrics. Compared to the IMF method, which also demonstrates excellent performance, our bidirectional alignment strategy yielded significantly better fusion results, outperforming IMF's unidirectional alignment strategy. 

 \begin{figure}[!t] \centering
	\subfigure[Visual comparison of MDF-FR ablation experiments]  {\includegraphics[height=1.3in,width=3.0in,angle=0]{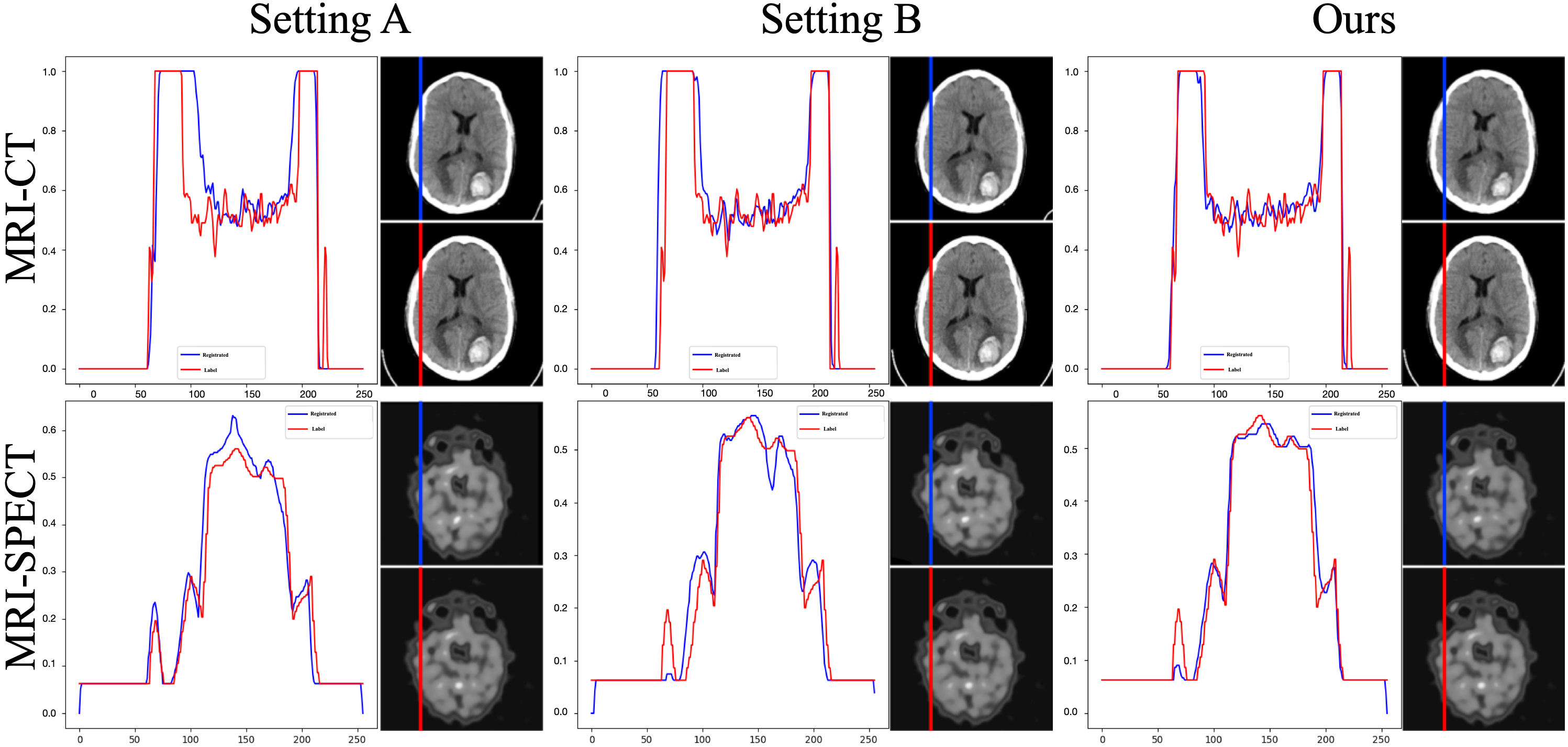}}
	\subfigure[Visual comparison of MDF-FR ablation experiments]   {\includegraphics[height=1.1in,width=3.0in,angle=0]{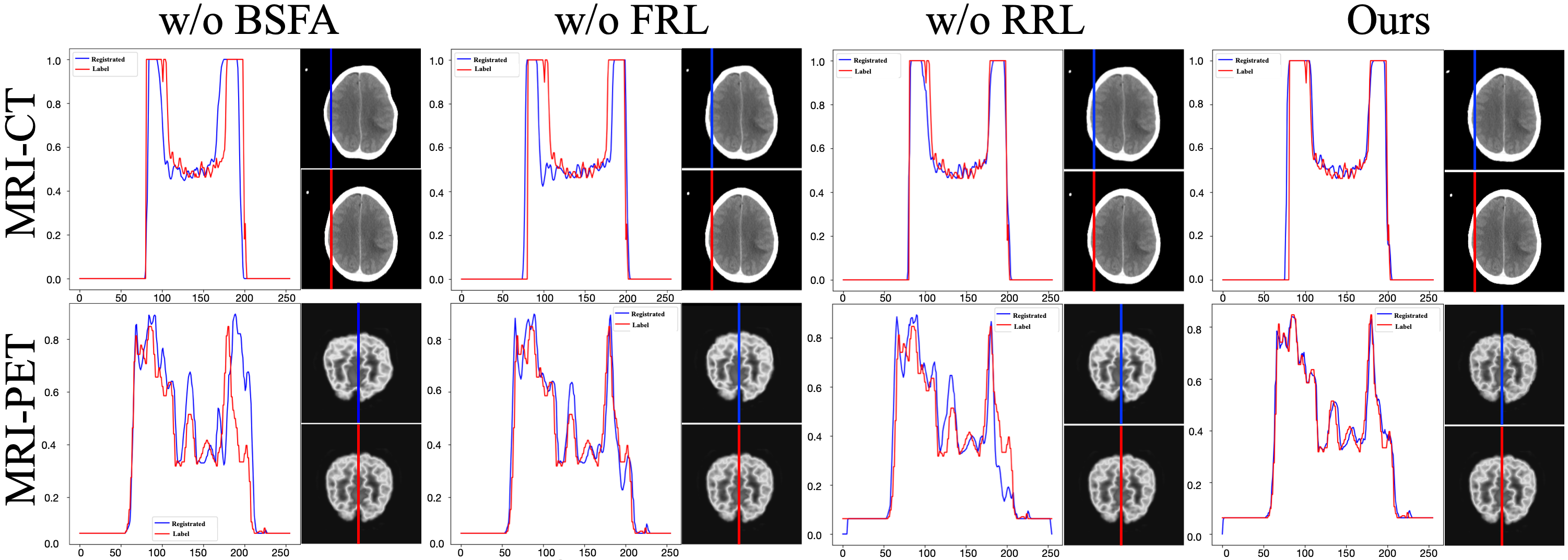}}
	\caption{Ablation experiments of MDF-FR and BSFA.}\label{fig6}
\end{figure}

\subsection{Ablation Study}
\textbf{Effectiveness of MDF-FR}: To verify the effectiveness of MDF-FR, we designed Setting A and Setting B. In Setting A, neither MFRH swapping nor the ${{\cal L}_{ce2}} $ was used.
In contrast, Setting B did not involve MFRH swapping but did use the ${{\cal L}_{ce2}} $.
The experimental results, as shown in Fig. \ref{fig6}(a), indicate that the proposed method achieves better alignment and fusion performance when MDF-FR is included. 
\textbf{Effectiveness of BSFA}: 
To verify the effectiveness of BSFA, we conducted several experiments. First, we removed BSFA entirely to assess its impact on overall alignment performance and fusion results. Next, we removed the FRL from BSFA, retaining only the RRL. Finally, we kept only the FRL and removed the RRL. The alignment results shown in Fig. \ref{fig6}(b) indicate that the proposed method achieves excellent alignment only when BSFA is fully implemented. 

\section{Conclusion}
This paper presents a single-stage multimodal medical image registration and fusion framework. Unlike traditional two-stage methods, it reduces model complexity with a shared feature encoder. By incorporating MDF-FR, the framework addresses modal differences in cross-modal feature alignment. The MFRH for each input integrates global image features, retaining complementary information across modalities. Additionally, a bidirectional stepwise alignment strategy predicts deformation fields using vector displacement principles. The method preserves fused information's integrity and diversity and shows potential for clinical applications requiring precise and efficient registration and fusion. 

\section{Acknowledgments}
This work was supported in part by the National Science Foundation of China under Grant 62161015, Grant 62471448, and Grant 62102338; in part by the Yunnan Fundamental Research Projects under Grant 202301AV070004; in part by  Shandong Provincial Natural Science Foundation under Grant ZR2024YQ004; and in part by TaiShan Scholars Youth Expert Program of Shandong Province under Grant No.tsqn202312109.

\bibliography{aaai25}

\twocolumn[%
\begin{center}
\vspace{10pt}

{\fontsize{16}{24}\selectfont \textbf{Supplementary Materials for ``BSAFusion: A Bidirectional Stepwise Feature Alignment Network for Unaligned Medical Image Fusion''}} 

\vspace{10pt}

{\large Huafeng Li$^{1}$, Dayong Su$^{1}$, Qing Cai$^{2*}$, Yafei Zhang$^{1*}$}

\vspace{6pt}

{\footnotesize 
$^{1}$School of Information Engineering and Automation, Kunming University of Science and Technology, \\Kunming 650500, China \\
$^{2}$School of Information Science and Engineering, Ocean University of China, Qingdao 266100, China \\[3pt]
\texttt{hfchina99@163.com, dayongsu@outlook.com, cq@ouc.edu.cn, zyfeimail@163.com}}
\end{center}

\vspace{6pt}
]

\section{Supplementary Experiments}

\subsection{1. Comparison with Registration+Fusion}
\begin{figure}[!htpb]
	\begin{center}
		\includegraphics[width=0.9\linewidth]{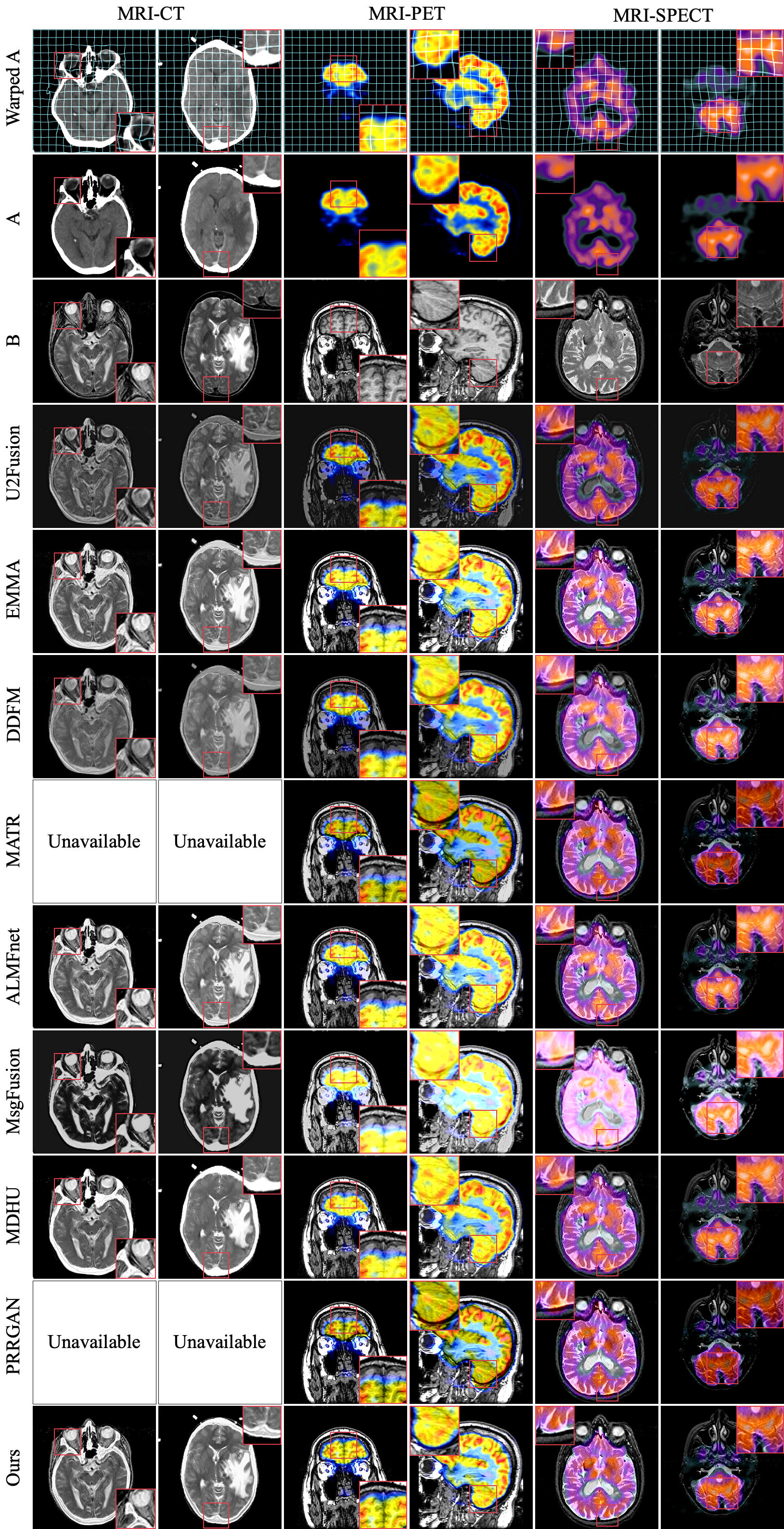}
	\end{center}
	\caption{Visual comparison of fusion results between the Registration+Fusion method and our proposed method. The first row displays the image to be fused after applying deformation, the second row shows the label corresponding to the first column image, and the third row presents the MRI image to be fused. Rows 4 to 9 exhibit the results obtained by different fusion methods.}
	\label{s1}
\end{figure}
\begin{figure*}[h!]
	\centering
	\includegraphics[width=0.9\textwidth]{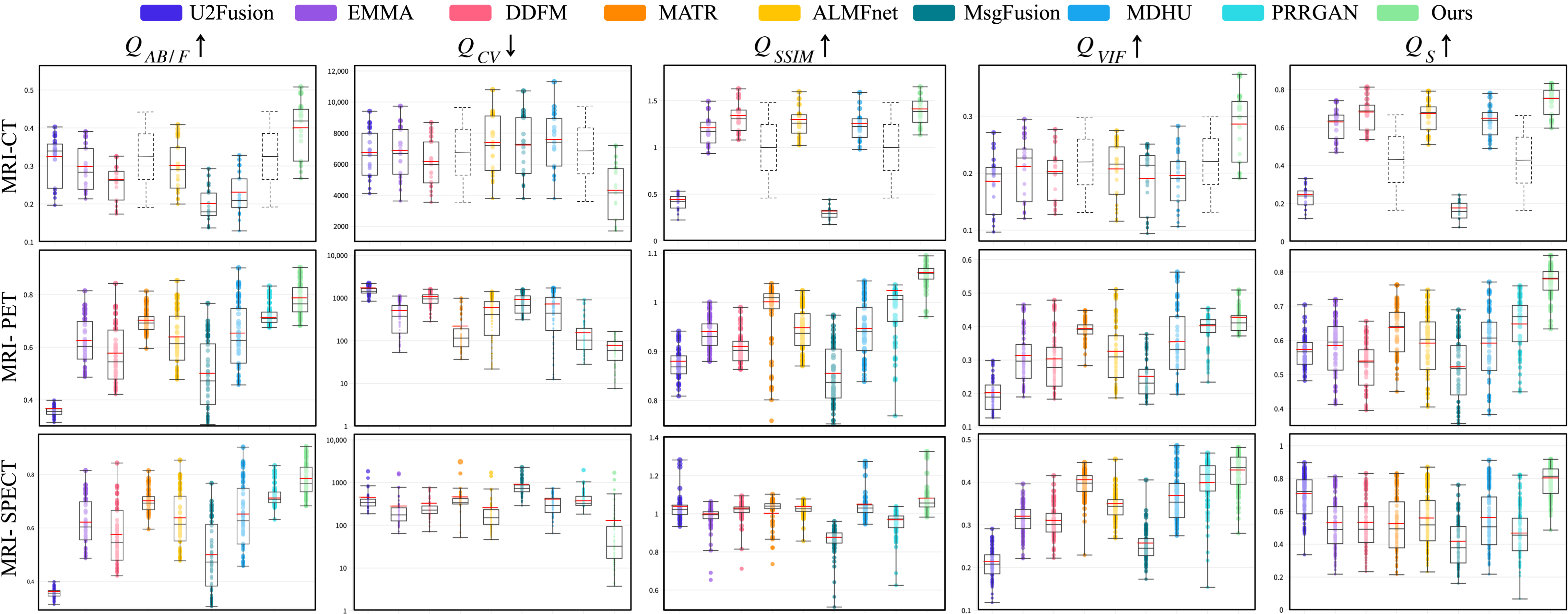}
	\caption{Comparison of objective evaluation results for fusion outcomes: Registration+Fusion Method vs. the Proposed Method. The x-axis represents each method, while the y-axis indicates the evaluation metric values. The upper and lower boundaries of each box denote the upper and lower quartiles, respectively. The black line represents the median, and the red line indicates the mean. A dashed box in the MRI-CT comparison indicates that the fusion result for that method is unavailable.}
	\label{s2}
\end{figure*}

\thispagestyle{firstpage} 
In the main text, we only compared the proposed method with the Joint Registration and Fusion method in terms of visual effects and objective data. To further verify the effectiveness of our method, we compare it with the Registration+Fusion approach in this supplementary material. For the Registration+Fusion approach, we selected IMSE \cite{48}, a high-performance registration method, to accurately align the images. Subsequently, we employed eight widely adopted image fusion methods, namely U2Fusion \cite{25}, EMMA \cite{42}, DDFM \cite{43}, MATR \cite{44}, ALMFnet \cite{45}, MsgFusion \cite{46}, PRRGAN \cite{47}, and MDHU \cite{48}, to fuse them. Among these methods, the first three are specifically designed for multimodal image fusion, while the remaining ones focus on multimodal medical image fusion. Note that the MATR method, which primarily targets the fusion of anatomical and functional images and emphasizes the preservation of functional information in PET and SPECT, was not designed for the effective fusion of MRI and CT images. Therefore, the MRI-CT fusion results using MATR are missing in Figure \ref{s1}. In addition, as we are unable to obtain the parameters of the PRRGAN model for the fusion of MRI-CT image pairs, we are also unable to display the fusion results for MRI-CT in Figure \ref{s2}. Based on the fusion results presented in Figure \ref{s1}, it is evident that the proposed method not only demonstrates stronger structural and texture preservation capabilities but also minimizes distortion and artifacts caused by feature misalignment. In contrast, traditional methods often produce suboptimal fusion results due to the separate registration and fusion steps, which prevent the registration process from being adjusted based on the fusion quality. Our method effectively addresses this issue, leading to higher-quality fusion results. To provide a direct and clear comparison of the metric differences among various methods, Table  \ref{table1},  \ref{table2},  \ref{table3}, present the quantitative comparisons of our method with single-stage methods and two-stage methods.

To more intuitively demonstrate the advantages of our proposed method, we present a box plot to quantitatively compare it with traditional methods. As shown in Figure. \ref{s2}, our method achieves the best average performance among the compared methods across three cross-modal scenarios and five evaluation metrics. Overall, while traditional methods can handle unaligned image fusion, their inherent disadvantages and unstable performance limit their wider application. Our proposed method combines registration and fusion, achieving both efficiency and stability, thereby offering broader applicability.

\begin{table}[h]\small
\centering
\caption{Comparison on MRI-CT dataset.}
\begin{adjustbox}{width=\columnwidth,center}
\begin{tabular}{|c |c| c| c| c| c|}
\hline
\textbf{Methods} & ${Q_{AB/F} \uparrow}$ & ${Q_{CV} \downarrow}$ & ${Q_{SSIM} \uparrow}$ & ${Q_{VIF} \uparrow}$ & ${Q_{S} \uparrow}$ \\ \hline
U2Fusion & 0.3170 & 6580.8 & 0.4093 & 0.1813 & 0.2322 \\
EMMA & 0.2911 & 6695.8 & 1.1812 & 0.2067 & 0.6198 \\
DDFM & 0.2557 & 5981.4 & 1.3125 & 0.1973 & 0.6727 \\
MATR & - & - & - & - & - \\
ALMFnet & 0.2934 & 7200.5 & 1.2659 & 0.2026 & 0.6594 \\
MsgFusion & 0.1936 & 7090.4 & 0.2908 & 0.1860 & 0.1616 \\
MDHU & 0.2237 & 7417.6 & 1.2268 & 0.1906 & 0.6330 \\
PRRGAN & - & - & - & - & - \\
UMF-CMGR & 0.2501 & 4638.7 & 1.3460 & 0.2363 & 0.6924 \\
SuperFusion & 0.2767 & 4828.9 & 0.9746 & 0.2300 & 0.5200 \\
MURF & 0.3298 & 5554.6 & 1.2679 & 0.2105 & 0.6717 \\
IMF & 0.2665 & 4439.6 & 1.3425 & 0.2413 & 0.6906 \\
PAMRFuse & 0.0879 & 5408.0 & 0.1820 & 0.1012 & 0.0881 \\
Ours & \textbf{0.3927} & \textbf{4155.1} & \textbf{1.3838} & \textbf{0.2811} & \textbf{0.7390} \\ \hline
\end{tabular}\label{table1}
\end{adjustbox}
\end{table}

\subsection{2. Further Discussion}
\begin{figure*}[t!]
	\centering
	\includegraphics[width=0.92\textwidth]{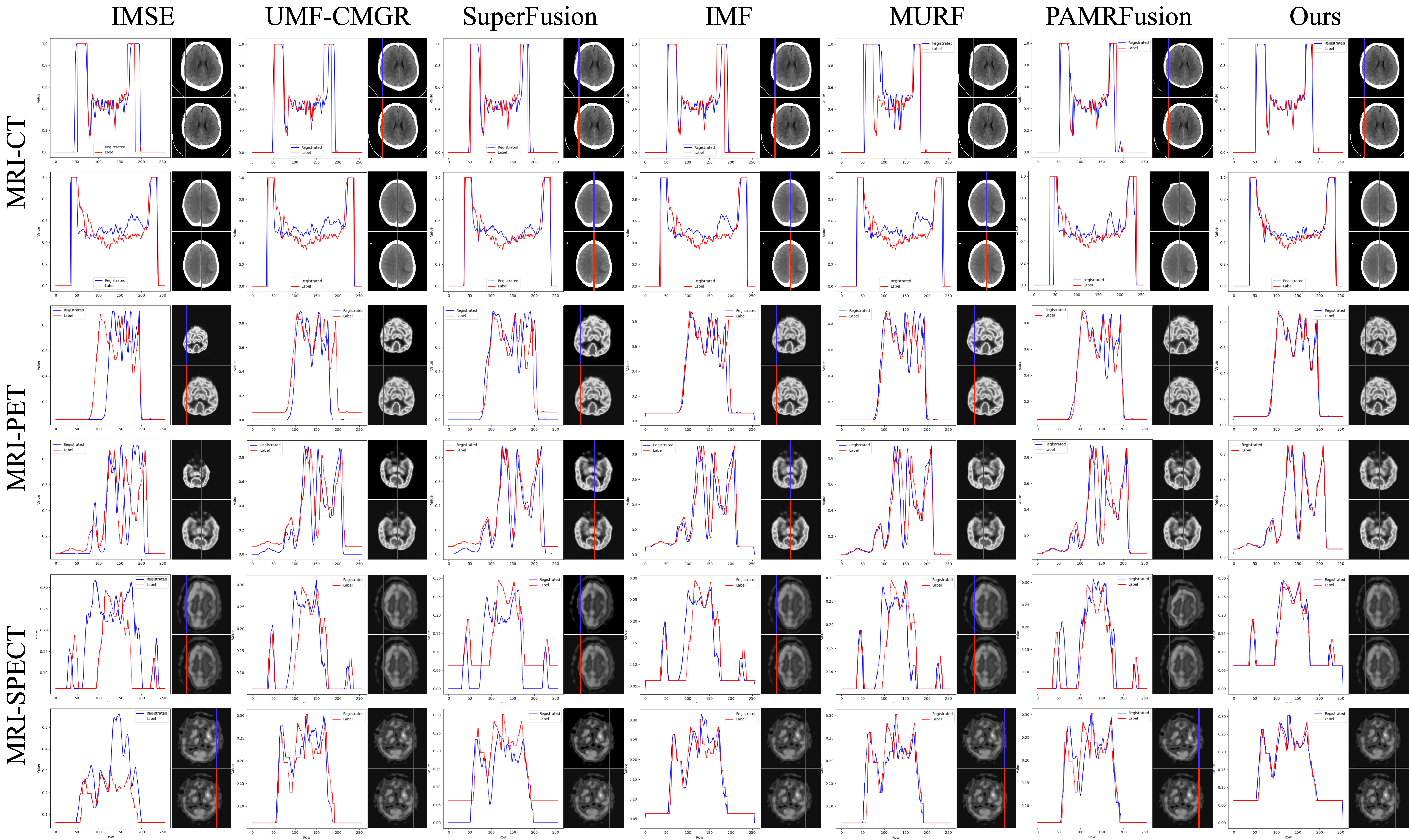}
	\caption{Visualization of feature alignment effect.}
	\label{s3}
\end{figure*}

Due to visual limitations, comparing the alignment effects of the images in Figure \ref{s1} and Figure \ref{s2} is challenging. To address this, we visualized the alignment effects of different methods in Figure \ref{s3}. In this figure, the blue and red curves represent the overlap of pixels at the corresponding blue and red line positions in the aligned image and the ground truth image, respectively. Since the blue line in the registered image should align with the red line in the ground truth image, greater registration accuracy results in a higher overlap between the two curves. The figure clearly shows that our method achieves superior alignment across various modal scenarios, particularly with complex structures and significant modal differences, such as in MRI-PET images, where its advantages are most pronounced.

\begin{table}[h]\small
\centering
\caption{Comparison MRI-PET dataset.}
\begin{adjustbox}{width=\columnwidth,center}
\begin{tabular}{|c |c| c| c| c| c|}
\hline
\textbf{Methods} & ${Q_{AB/F} \uparrow}$ & ${Q_{CV} \downarrow}$ & ${Q_{SSIM} \uparrow}$ & ${Q_{VIF} \uparrow}$ & ${Q_{S} \uparrow}$ \\ \hline
U2Fusion & 0.3558 & 1483.5 & 0.8739 & 0.1922 & 0.5650 \\
EMMA & 0.6139 & 440.7 & 0.9345 & 0.3037 & 0.5766 \\
DDFM & 0.5670 & 946.7 & 0.9048 & 0.2937 & 0.5296 \\
MATR & 0.6107 & 1124.1 & 0.9084 & 0.3525 & 0.5355 \\
ALMFnet & 0.6286 & 507.4 & 0.9426 & 0.3167 & 0.5837 \\
MsgFusion & 0.4906 & 786.4 & 0.8494 & 0.2425 & 0.5127 \\
MDHU & 0.6430 & 616.6 & 0.9414 & 0.3452 & 0.5835 \\
PRRGAN & 0.5989 & 1284.4 & 0.8889 & 0.3621 & 0.5273 \\
UMF-CMGR & 0.6376 & 206.2 & 0.9302 & 0.3197 & 0.6536 \\
SuperFusion & 0.5055 & 545.1 & 0.8787 & 0.2763 & 0.5764 \\
MURF & 0.5413 & 578.1 & 0.8684 & 0.2601 & 0.6420 \\
IMF & 0.6242 & 265.9 & 0.9136 & 0.3084 & 0.5458 \\
PAMRFuse & 0.0908 & 3050.6 & 0.4252 & 0.1762 & 0.2723 \\
Ours & \textbf{0.7770} & \textbf{67.7} & \textbf{1.0549} & \textbf{0.4189} & \textbf{0.7726} \\ \hline
\end{tabular}\label{table2}
\end{adjustbox}
\end{table}

\begin{table}[h]\small
\centering
\caption{Comparison on MRI-SPECT dataset.}
\begin{adjustbox}{width=\columnwidth,center}
\begin{tabular}{|c |c| c| c| c| c|}
\hline
\textbf{Methods} & ${Q_{AB/F} \uparrow}$ & ${Q_{CV} \downarrow}$ & ${Q_{SSIM} \uparrow}$ & ${Q_{VIF} \uparrow}$ & ${Q_{S} \uparrow}$ \\ \hline
U2Fusion & 0.4340 & 392.2 & 1.0268 & 0.2078 & 0.6911 \\
EMMA & 0.6177 & 240.3 & 0.9812 & 0.3127 & 0.5135 \\
DDFM & 0.6086 & 279.7 & 1.0144 & 0.3050 & 0.5156 \\
MATR & 0.6270 & 498.7 & 0.9856 & 0.3993 & 0.5055 \\
ALMFnet & 0.6463 & 219.9 & 1.0234 & 0.3427 & 0.5413 \\
MsgFusion & 0.3657 & 790.4 & 0.8608 & 0.2515 & 0.3984 \\
MDHU & 0.6222 & 354.1 & 1.0354 & 0.3619 & 0.5440 \\
PRRGAN & 0.5775 & 819.0 & 0.9052 & 0.3614 & 0.4492 \\
UMF-CMGR & 0.4961 & 292.9 & 1.0551 & 0.3333 & 0.5179 \\
SuperFusion & 0.4989 & 336.0 & 0.9695 & 0.2988 & 0.5650 \\
MURF & 0.4873 & 456.1 & 0.8226 & 0.2707 & 0.5870 \\
IMF & 0.5109 & 295.1 & 0.9997 & 0.3105 & 0.4776 \\
PAMRFuse & 0.1278 & 1557.5 & 0.5596 & 0.1408 & 0.3608 \\
Ours & \textbf{0.7679} & \textbf{110.8} & \textbf{1.0661} & \textbf{0.4222} & \textbf{0.7851} \\ \hline
\end{tabular}\label{table3}
\end{adjustbox}
\end{table}

\subsection{3. Ablation Study}
In the main text, we conducted ablation experiments to investigate the impact of the model's core components, MDF-FR and BSFA, on feature alignment performance. In this supplementary material, we further evaluate the impact of MDF-FR, BSFA, and MMFF on fusion performance.

\textbf{Effectiveness of MDF-FR in Fusion Performance}: The main purpose of MDF-FR is to reduce the negative impact of modality differences on feature alignment, and the effectiveness of feature alignment is directly related to the quality of fusion performance. To further validate the effectiveness of MDF-FR, we evaluated its impact on fusion performance under two configurations, Setting A and Setting B, as constructed in the main text ablation experiments. As shown in Table \ref{table4}, regardless of the specific multimodal medical image fusion task, the model only exhibits the best performance on all evaluation metrics when MDF-FR is fully included, which fully demonstrates the effectiveness of MDF-FR. 

\begin{table}[h]
	\centering\small
	\caption{MDF-FR Ablation Study}
	\renewcommand\arraystretch{1.3}
	\fontsize{10}{10}\selectfont 
	\begin{adjustbox}{width=\columnwidth,center}
	\begin{tabular}{|c|c|c|c|c|c|c|}
		\hline
		\multirow{2}{*}{Datasets} & \multirow{2}{*}{Settings} & \multicolumn{5}{c|}{Metrics} \\
		\cline{3-7}
		& & $Q_{AB/F} \uparrow$ & $Q_{CV} \downarrow$ & $Q_{SSIM} \uparrow$ & $Q_{VIF} \uparrow$ & $Q_{S} \uparrow$ \\
		\hline
		\multirow{3}{*}{MRI-CT} & Setting A & 0.3453 & \textbf{4120.0} & 1.3681 & 0.2638 & 0.7191 \\
		
		& Setting B & 0.3544 & 4912.7 & 1.3751 & 0.2561 & 0.7074 \\
		
		& Ours      & \textbf{0.3927} & 4155.1 & \textbf{1.3838} & \textbf{0.2811} & \textbf{0.7390} \\
		\hline
		\multirow{3}{*}{MRI-PET} & Setting A & 0.7282 & 99.2 & 1.0419 & 0.4027 & 0.7567 \\
		
		& Setting B & 0.7216 & 101.9 & 1.0270 & 0.3941 & 0.7466 \\
		
		& Ours      & \textbf{0.7770} & \textbf{67.7} & \textbf{1.0549} & \textbf{0.4189} & \textbf{0.7726} \\
		\hline
		\multirow{3}{*}{MRI-SPECT} & Setting A & 0.7275 & 167.7 & 1.0535 & 0.3586 & 0.7586 \\
		
		& Setting B & 0.7408 & 138.5 & 1.0422 & 0.4073 & 0.7670 \\
		
		& Ours      & \textbf{0.7678} & \textbf{110.8} & \textbf{1.0661} & \textbf{0.4222} & \textbf{0.7851} \\
		\hline
	\end{tabular}\label{table4}
	\end{adjustbox}
\end{table}

\textbf{Effectiveness of BSFA in Fusion Performance}: BSFA is mainly used to predict the correspondence between features and obtain a deformation field that reflects the deformation between two images. Therefore, BSFA's performance is directly linked to the quality of the fusion outcomes. Here, we conduct an ablation experiment on BSFA following the settings in the main text to evaluate its impact based on the objective quality of the fusion results. From Table \ref{table5}, it is evident that the fusion performance of the model decreases when either RRL or FRL is absent. This is because the lack of RRL or FRL directly reduces the accuracy of deformation field prediction, thereby affecting fusion performance. These results highlight the importance of RRL and FRL in BSFA, underscoring its effectiveness.

\begin{table}[h]
	\centering\small
	\caption{BSFA Ablation Study. ``w/o R" indicates the absence of RRL, ``w/o F" indicates the absence of FRL, and ``Bid" indicates the presence of both RRL and FRL.}
	\renewcommand\arraystretch{1.3}
	\fontsize{10}{10}\selectfont 
	\begin{adjustbox}{width=\columnwidth,center}
	\begin{tabular}{|c|c|c|c|c|c|c|}
		\hline
		\multirow{2}{*}{Datasets} & \multicolumn{1}{c|}{Alignment} & \multicolumn{5}{c|}{Metrics} \\
		\cline{3-7}
		& \multicolumn{1}{c|}{Mode} & $Q_{AB/F} \uparrow$ & $Q_{CV} \downarrow$ & $Q_{SSIM} \uparrow$ & $Q_{VIF} \uparrow$ & $Q_{S} \uparrow$ \\
		\hline
		\multirow{4}{*}{MRI-CT} & w/o BSFA & 0.3279 & 5336.1 & 1.3418 & 0.2443 & 0.7045 \\
		
		& w/o R & 0.3753 & 4368.0 & 1.3683 & 0.2661 & 0.7274 \\
		
		& w/o F & 0.3464 & 4517.9 & 1.3681 & 0.2570 & 0.7238 \\
		
		& Bid & \textbf{0.3927} & \textbf{4155.1} & \textbf{1.3838} & \textbf{0.2810} & \textbf{0.7390} \\
		\hline
		\multirow{4}{*}{MRI-PET}& w/o BSFA & 0.7236 & 116.4 & 1.0244 & 0.3878 & 0.7453 \\
		& w/o R & 0.6817 & 217.7 & 1.0260 & 0.3613 & 0.7419 \\
		
		& w/o F & 0.7272 & 101.3 & 1.0381 & 0.3995 & 0.7541 \\
		
		& Bid & \textbf{0.7770} & \textbf{67.7} & \textbf{1.0549} & \textbf{0.4189} & \textbf{0.7726} \\
		\hline
		\multirow{4}{*}{MRI-SPECT}& w/o BSFA & 0.7277 & 144.3 & 1.0563 & 0.3746 & 0.7716 \\
		& w/o R & 0.7298 & 152.8 & 1.0590 & 0.3779 & 0.7731 \\
		
		& w/o F & 0.7259 & 142.9 & 1.0566 & 0.3766 & 0.7712 \\
		
		& Bid & \textbf{0.7678} & \textbf{110.8} & \textbf{1.0661} & \textbf{0.4222} & \textbf{0.7851} \\
		\hline
	\end{tabular}\label{table5}
	\end{adjustbox}
\end{table}

\textbf{Effectiveness of Stepwise Alignment and FusionBLK}: To verify the effectiveness of combining multiple FRL\&RRL blocks and FusionBLK modules, we analyzed the fusion performance in the absence of MMFF and examined how the performance changes with different numbers of FRL\&RRL blocks and FusionBLK modules. The experimental results, presented in Table \ref{table6}, show that when MMFF is omitted and fusion is achieved through feature concatenation, the fusion performance decreases. This indicates that MMFF is more effective at fusing image features than direct feature concatenation. Additionally, the data in Table \ref{table6} reveal that the fusion performance reaches its optimal level when both the numbers of FRL\&RRL blocks and FusionBLK modules are set to 5.

\begin{table}[h]
	\centering\small
	\caption{MMF Performance Evaluation for FRL\&RRL, and FusionBLK with Different Metrics. ``N-FR'' and ``N-FBL'' denote the number of ``FR\&RRL'' and FusionBLK respectively. ``-'' indicates that the data is not given.}
	\renewcommand\arraystretch{1.3}
	\fontsize{10}{10}\selectfont 
	\begin{adjustbox}{width=\columnwidth,center}
		\begin{tabular}{|c|c|c|c|c|c|c|}
			\hline
			Metrics & \diagbox[dir=NW]{N-FR}{N-FBL} &0 & 3 &4 &5 &6 \\
			\hline
			\multirow{4}{*}{$Q_{AB/F} \uparrow$} & 3&- & 0.6406 & 0.6508 & 0.6666 &0.6849 \\
			
			& 4&- & 0.6694 & 0.6405 & 0.6457 &0.6730 \\
			
			& 5&0.6765 & 0.6704 & 0.6982 & \textbf{0.7218}&0.7134 \\
			& 6&- & 0.6803 & 0.6930 &0.7098 &0.7190 \\
			\hline
			\multirow{4}{*}{$Q_{CV} \downarrow$} & 3 &-& 848.5 & 846.2 & 835.7 &832.5 \\
			
			& 4 &- & 838.1 & 830.2 & 828.2 &823.1 \\
			
			& 5 &758.0 & 788.9 & 783.5 & \textbf{627.4} &716.6 \\
			& 6 &- & 748.2 & 730.6 &701.9 &683.6 \\
			\hline
			\multirow{4}{*}{$Q_{SSIM} \uparrow$} & 3 &- & 1.0759 & 1.0753 & 1.0817 &1.0825 \\
			
			& 4 &- & 1.0805 & 1.0816 & 1.0818 &1.0918 \\
			
			& 5 &1.0867 & 1.0911 & 1.0896 & \textbf{1.1039} &1.0936 \\
			
			& 6 &- & 1.0934 & 1.0942 &1.0897 &1.0965 \\
			\hline
		\end{tabular}\label{table6}
	\end{adjustbox}
\end{table}

\subsection{4. Analysis of Model Complexity and Limitations}
\begin{figure}[t!]
	\centering
	\includegraphics[width=0.4\textwidth]{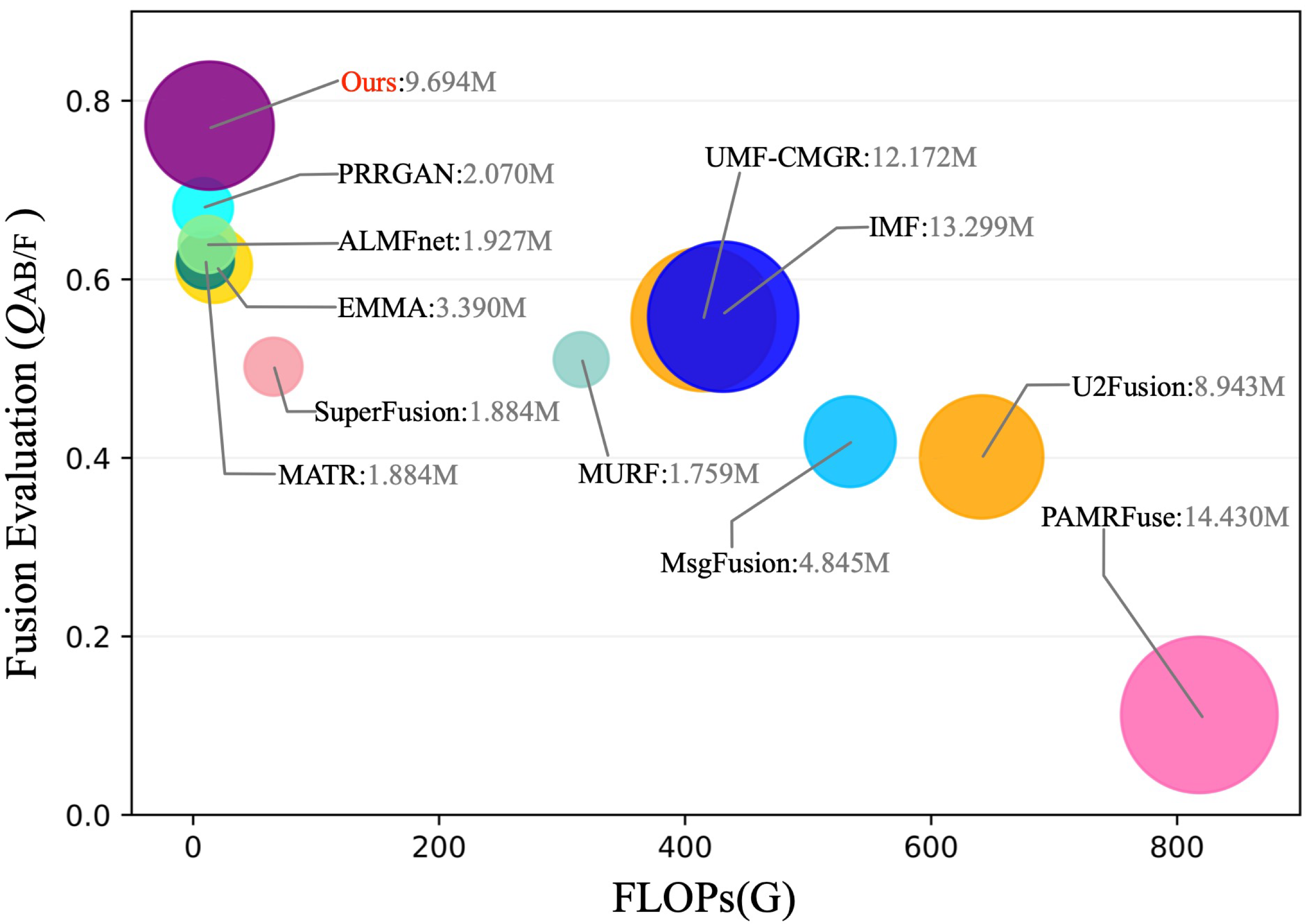}
	\caption{Model complexity analysis. The x-axis in the figure represents the FLOPs (in billions) obtained by the model on a 256$\times$256 image input, the y-axis represents the mean $Q_{AB/F}$ score, and the bubble size indicates the number of parameters.}
	\label{s4}
\end{figure}

Fig.\ref{s4} presents the complexity analysis results of our method in comparison to existing approaches. The results show that our proposed method not only achieves superior performance in image fusion tasks but also has lower FLOPs compared to most other methods, indicating its relatively low model complexity. This is largely attributed to the adoption of an integrated single-stage design, which enhances overall performance. However, it is also noted that our method has a relatively high parameter scale, though it remains moderate compared to other methods. This is primarily due to the use of Transformer layers for feature extraction and the incorporation of multiple FRL\&RRL and FusionBLK modules for feature alignment and fusion, which increase the model's parameter scale.

\subsection{5. Limitations and Future Work}
Although this article adopts the single-stage integrated model design, avoiding the use of multiple encoders, the model is still not sufficiently lightweight due to the incorporation of multiple FRL\&RRL and FusionBLK modules. Therefore, in future work, we plan to further optimize and expand this framework, aiming to make it more lightweight while also being applicable to a wider range of medical imaging modes and application scenarios.

\bibliography{aaai25}

\end{document}